\documentclass[draft]{amsart}
\usepackage{amssymb,amscd}

\begin{document}

\title[Commuting differential operators]{Self-adjoint commuting differential operators and commutative subalgebras of the Weyl algebra}
\author{{Andrey E. Mironov}}\thanks{This paper was finished in the Hausdorff Research Institute for Mathematics (Bonn). The author is grateful to the Institute
for hospitality. This work was also partially supported by the Presidium of the Russian Academy of Sciences (under
the program "Fundamental Problems of Nonlinear Dynamics"); grant MD-5134.2012.1 from the
President of Russia; and a grant from Dmitri Zimin's "Dynasty" foundation.}
\address{Sobolev Institute of Mathematics, Novosibirsk, Russia,
\quad
\emph
\newline\indent Novosibirsk State University,
\quad
\emph{and}
\newline\indent Laboratory of Geometric Methods in Mathematical Physics, Moscow State University}
\email{mironov@math.nsc.ru}

\maketitle

\begin{quote}
\noindent{\sc Abstract. } In this paper we study self-adjoint
commuting ordinary differential operators.  We find
sufficient conditions when an operator of fourth order commuting
with an operator of order $4g+2$ is self-adjoint. We introduce an equation on
potentials $V(x),W(x)$ of the self-adjoint
operator $L=(\partial_x^2+V)^2+W$ and some additional data. With the help of this equation we
find the first example of commuting differential operators of rank
two corresponding to a spectral curve of arbitrary genus. These
operators have polynomial coefficients and define commutative
subalgebras of the first Weyl algebra.
 \medskip
 \end{quote}

\section{Introduction}
 The problem of finding commuting differential operators is a classical problem of differential
equations (for the first results see \cite{W}--\cite{BC}). In the case of operators of rank greater than one, this problem has not been solved until now.
In this paper we study self-adjoint commuting ordinary differential operators. One of the main results of this paper is the following. We find
an example of commuting differential operators of rank two corresponding to spectral curves of arbitrary genus.

If two differential operators
$$
L_n=\partial_x^n+\sum_{i=0}^{n-2}u_{i}(x)\partial_x^i,\
\
L_m=\partial_x^m+\sum_{i=0}^{m-2}v_{i}(x)\partial_x^{i}\
$$
commute, then there is a nonzero polynomial $R(z,w)$ such that $R(L_n,L_m)=0$ (see \cite{BC}).
The curve $\Gamma$ defined by $R(z,w)=0$ is called the {\it spectral curve.}
This curve  parametrizes common eigenvalues of the operators.  If
$$
 L_n\psi=z\psi,\qquad L_m\psi=w\psi,
$$
then $(z,w)\in\Gamma$.  For almost all $(z,w)\in\Gamma$ the dimension of the space of common eigenfunctions
$\psi$ is the same. The dimension is called the {\it rank}. The rank equals the greatest common divisor of $m$ and $n$.

In this paper we consider only commuting ordinary differential operators whose spectral curves are smooth. Commutative rings of such operators were
classified by Kriche\-ver \cite{K1}, \cite{K}. The ring is
determined by the spectral curve and some additional spectral data.
If the rank is one, then the spectral data define  commuting operators
by explicit formulas (see \cite{K1}). In the case of operators of rank greater than one there are the following results.
Krichever and Novikov \cite{KN1}, \cite{KN2} using the method of deformation of Tyurin parameters
found operators of rank two corresponding to an elliptic spectral curves. These operators were studied in the papers \cite{GN}--\cite{Deh}.
Mokhov \cite{Mokh}, using the same method found operators of rank three also corresponding
to elliptic spectral curves. Besides this there are examples of operators of rank grater than one corresponding to spectral curves of genus
$2,3$ and $4$ (see \cite{M1}--\cite{Z}).

The main results of this paper are the following. We consider a pair $L_4,L_{4g+2}$ of commuting differential operators of rank two
whose spectral curve is a hyperelliptic curve $\Gamma$ of genus $g$
\begin{equation}\label{v1}
 w^2=F_g(z)=z^{2g+1}+c_{2g}z^{2g}+\dots+c_0.
\end{equation}
Operators $L_4$ and $L_{4g+2}$ satisfy the equation $(L_{4g+2})^2=F_g(L_4)$.
The curve $\Gamma$ has a holomorphic involution
$$
 \sigma:\Gamma\rightarrow\Gamma,\qquad \sigma(z,w)=(z,-w).
$$
Common eigenfunctions of $L_4$ and $L_{4g+2}$ satisfy the second order differential equation \cite{K}
\begin{equation}\label{u1}
 \psi''(x,P)=\chi_1(x,P)\psi'(x,P)+\chi_0(x,P)\psi(x,P).
\end{equation}
The coefficients $\chi_0(x,P),\chi_1(x,P)$ are rational functions on $\Gamma$ with $2g$ simple poles depending on $x$, $\chi_0$ has also an
additional simple pole at infinity. These functions satisfy Krichever's equations
(see below). To find operators $L_4,L_{4g+2}$ it is enough to find $\chi_0,\chi_1$.

It is not difficult to prove
that if $\chi_1$ is invariant under the involution $\sigma$, then the operator $L_4$ is
self-adjoint. S.P. Novikov has proposed the conjecture that the inverse is also true. In this paper we prove this conjecture.

\vspace{0.4cm}

\noindent{\bf Theorem 1} {\it The operator $L_4$ is self-adjoint if and only if
\begin{equation}\label{u2}
 \chi_1(x,P)=\chi_1(x,\sigma(P)).
\end{equation}
}
\vspace{0.4cm}

At $g=1$ Theorem 1 was proved by Grinevich and Novikov \cite{GN}.

Let us assume that the operator $L_4$ is self-adjoint
$$
 L_4=(\partial_x^2+V(x))^2+W(x),
$$
then the functions $\chi_0,\chi_1$ have simple poles at some points
$$(\gamma_i(x),\pm\sqrt{F_g(\gamma_i(x))}),\ 1\leq i\leq g.$$
In the next theorem we find the form of $\chi_0(x,P),\chi_1(x,P).$

\vspace{0.4cm}

\noindent{\bf Theorem 2} {\it
If operator $L_4$ is self-adjoint, then
$$
 \chi_0=-\frac{1}{2}\frac{Q''}{Q}+\frac{w}{Q}-V, \qquad \chi_1=\frac{Q'}{Q},
$$
where $Q=(z-\gamma_1(x))\dots(z-\gamma_g(x))$. Functions $Q,V,W$ satisfy the equation
\begin{equation}\label{e1}
 4F_g(z)=4(z-W)Q^2-4V(Q')^2+(Q'')^2-2Q'Q^{(3)}
 +2Q(2V'Q'+4VQ''+Q^{(4)}),
\end{equation}
where $Q',Q'',Q^{(k)}$ mean $\partial_xQ,\partial_x^2Q,\partial_x^kQ.$}

\vspace{0.4cm}

To find self-adjoint operators $L_4, L_{4g+2}$ it is enough to solve the equation (\ref{e1}).

In this paper we find partial solutions of the equation  for arbitrary $g$. These solutions correspond to operators with polynomial coefficients.

\vspace{0.4cm}

\noindent{\bf Theorem 3} {\it
The operator
$$
 L^{^{\sharp}}_4=(\partial_x^2+\alpha_3x^3+\alpha_2 x^2+\alpha_1x+\alpha_0)^2+g(g+1)\alpha_3x, \qquad \alpha_3\ne 0
$$
commutes with a differential operator
$L_{4g+2}^{^{\sharp}}$ of order $4g+2$. The operators
$L_4^{^{\sharp}},$ $L_{4g+2}^{^{\sharp}}$ are operators of rank two.
For generic values of parameters $(\alpha_0,\alpha_1,\alpha_2,\alpha_3)$ the spectral
curve is a nonsingular hyperelliptic curve of genus $g$.
 }

\vspace{0.4cm}

If $g=1$, $\alpha_1=\alpha_2=0,\ \alpha_3=1$, then the operators $L^{^{\sharp}}_4,L_{4g+2}^{^{\sharp}}$ coincide with the famous
Dixmier operators \cite{D} whose spectral curve is an elliptic curve. Operators $L^{^{\sharp}}_4,L_{4g+2}^{^{\sharp}}$ define
commutative subalgebras in the first Weyl algebra $A_1$.
Theorem 3 means that the equation
$
 Y^2=X^{2g+1}+c_{2g}X^{2g}+\dots+c_0
$
has nonconstant solutions $X,Y\in A_1$ for some $c_i$.
It is easy to see that the group $ Aut(A_1)$ preserves the space of
all such solutions. It would be very interesting to describe the orbits of $ Aut(A_1) $
in the space of solutions under the action of $ Aut(A_1) $. This gives a chance to compare $End (A_1)$ and $Aut (A_1)$
(the Dixmier conjecture is: $End (A_1)= Aut (A_1)$).

In Section 2 we recall the method of deformations of Tyurin parameters.
In Sections 3--5 we prove Theorems 1--3.

The author is grateful to I.M. Krichever, O.I. Mokhov, S.P. Novikov and V.V.~So\-kolov
 for valuable discussions and stimulating interest.

\section{Operators of rank $l>1$}
 Common eigenfunctions of commuting differential operators are Baker--Akhiezer functions.
Let me recall the definition of the Baker--Akhiezer function at $l>1$ \cite{K}. We take the {\it spectral data}
$$
 \{\Gamma,q,k^{-1},\gamma,v,\omega(x)\},
$$
where $\Gamma$ is a Riemann surface of genus $g$,
$q$ is a fixed point on $\Gamma$, $k^{-1}$ is a local parameter near $q$,
$$
 \omega(x)=(\omega_0(x),\dots,\omega_{l-2}(x))
$$
is a set of smooth functions,  $\gamma=\gamma_1+\dots+\gamma_{lg}$ is a divisor on $\Gamma$, $v$ is a set of vectors
$$
 v_1,\dots,v_{lg},\qquad
v_i=(v_{i,1},\dots,v_{i,l-1}).
$$
The pair $(\gamma,v)$ is called the {\it Tyurin parameters}. The Tyurin parameters define a stable
holomorphic vector bundle on $\Gamma$ of rank $l$ and degree
$lg$ with holomorphic sections  $\eta_1,\dots,\eta_l$. The points  $\gamma_1,\dots,\gamma_{lg}$ are the points of the linear dependence
$$
 \eta_l(\gamma_i)=\sum_{i=1}^{l-1}v_{j,i}\eta_j(\gamma_i).
$$
The vector-function $\psi=(\psi_1,\dots,\psi_{l})$ is defined by the following properties.

1. In the neighbourhood of $q$ the vector-function $\psi$ has the form
$$
 \psi(x,P)=\left(\sum_{s=0}^{\infty}\xi_s(x)k^{-s}\right)\Psi_0(x,k),
$$
where $\xi_0=(1,0,\dots,0), \xi_i(x)=(\xi_i^1(x),\dots,\xi_i^l(x))$, the matrix  $\Psi_0$ satisfies the equation
$$
 \frac{d\Psi_0}{dx}=A\Psi_0,\
A=\left(
\begin{array}{cccccc}
 0 & 1 & 0 & \dots  & 0 & 0\\
 0 & 0 & 1 & \dots  & 0 & 0\\
 \dots & \dots & \dots & \dots &  \dots & \dots\\
 0 & 0 & 0 & \dots  & 0 & 1 \\
 k+\omega_0 & \omega_1 & \omega_2 & \dots & \omega_{l-2} & 0
 \end{array}\right).
$$
2.  The components of $\psi$ are meromorphic functions on $\Gamma\backslash\{q\}$ with the simple poles $\gamma_1,\dots,\gamma_{lg}$,
and
$$
 {\rm Res}_{\gamma_i}\psi_j=v_{i,j}{\rm Res}_{\gamma_i}\psi_{l},\quad 1\leq i\leq lg,\ 1\leq j\leq l-1.
$$
 For the rational function  $f(P)$ on $\Gamma$ with the unique pole of order $n$ at  $q$
there is a linear differential operator $L(f)$ of order $ln$ such that
$$
 L(f)\psi(x,P)=f(P)\psi(x,P).
$$
For two such functions $f(P),g(P)$ operators $L(f)$, $L(g)$ commute.

The main difficulty
to construct operators of rank $l>1$ is the fact that the Baker--Akhiezer function is not found explicitly.
But the operators can be found by the method of deformation of Tyurin parameters.

The common eigenfunctions of commuting differential operators of rank $l$ satisfy the linear differential equation of order $l$
$$
 \psi^{(l)}(x,P)=\chi_0(x,P)\psi(x,P)+\dots+\chi_{l-1}(x,P)\psi^{(l-1)}(x,P).
$$
Coefficients $\chi_i$ are rational functions on $\Gamma$ \cite{K} with simple poles
$P_1(x),\dots,P_{lg}(x)\in\Gamma$, and with the following expansions in the
neighbourhood of $q$
$$
\chi_0(x,P)=k+g_0(x)+O(k^{-1}),$$
$$
 \chi_j(x,P)=g_j(x)+O(k^{-1}), \ \
 0<j<l-1,
$$
$$
\chi_{l-1}(x,P)=O(k^{-1}).
$$
Let $k-\gamma_i(x)$ be a local parameter near $P_i(x)$. Then
$$
\chi_j=\frac{c_{i,j}(x)}{k-\gamma_i(x)}+d_{i,j}(x)+O(k-\gamma_i(x)).
$$
Functions $c_{ij}(x) ,d_{ij}(x)$ satisfy the following equations \cite{K}.

\vspace{0.4cm}

\noindent{\bf Theorem 4}
{\it
\begin{equation}\label{w1}
 c_{i,l-1}(x)=-\gamma'_i(x),
\end{equation}
\begin{equation}\label{w2}
 d_{i,0}(x)=v_{i,0}(x)v_{i,l-2}(x)+v_{i,0}(x)d_{i,l-1}(x)
 -v'_{i,0}(x),
\end{equation}
\begin{equation}\label{w3}
 d_{i,j}(x)=v_{i,j}(x)v_{i,l-2}(x)-v_{i,j-1}(x)+v_{i,j}(x)d_{i,l-1}(x)
 -v'_{i,j}(x), j\geq 1,
\end{equation}
where
$$
 v_{i,j}(x)=\frac{c_{i,j}(x)}{c_{i,l-1}(x)}, \ \ 0\leq j\leq l-1, \ 1\leq i\leq lg.
$$}

To find $\chi_i$ one should solve equations (\ref{w1})--(\ref{w3}).

\section{Proof of Theorem 1}

In the case of operators of rank two the common eigenfunctions of $L_4$ and $L_{4g+2}$ satisfy equation (\ref{u1}).
In the neighbourhood of $q$ we have the expansions
\begin{equation}\label{u3}
\chi_0=\frac{1}{k}+a_0(x)+a_1(x)k+O(k^2),\quad \chi_1=b_1(x)k+b_2(x)k^2+O(k^3).
\end{equation}
Functions $\chi_0,\chi_1$ have $2g$ simple poles $P_1(x),\dots,P_{2g}(x)$, and by Theorem 4

\begin{equation}\label{u4}
\chi_0(x,P)=\frac{-v_{i,0}(x)\gamma'_i(x)}{k-\gamma_i(x)}+d_{i,0}(x)+O(k-\gamma_i(x)),
\end{equation}
\begin{equation}\label{u5}
\chi_1(x,P)=\frac{-\gamma'_i(x)}{k-\gamma_i(x)}+d_{i,1}(x)+O(k-\gamma_i(x)),
\end{equation}
\begin{equation}\label{u6}
d_{i,0}(x)=v_{i,0}^2(x)+v_{i,0}(x)d_{i,1}(x)-v'_{i,0}(x).
\end{equation}
Let $\Gamma$ be the hyperelliptic spectral curve (\ref{v1}),
 $q=\infty\in \Gamma,$ $k=\frac{1}{\sqrt{z}}$.

Let us find coefficients of the operator of order 4 corresponding to  $z$, $L_4\psi=z\psi.$

\vspace{0.4cm}

\noindent{\bf Lemma 1} {\it The operator $L_4=\partial_x^4+f_2(x)\partial_x^2+f_1(x)\partial_x+f_0(x)$  has the following coefficients:
$$
 f_0=a_0^2-2a_1-2b_1'-a_0'',\qquad  f_1=-2(b_1+a_0'),\qquad f_2=-2a_0.
$$
Operator $L_4$ is self-adjoint if and only if $b_1=0$, herewith
$
 L_4=(\partial_x^2+V(x))^2+W(x),
$
where $V(x)=-a_0(x),$ $W=-2a_1(x)$.}

\vspace{0.4cm}
\noindent{\bf Proof.} From (\ref{u1}) it follows that the fourth derivative of $\psi$ is
$$
 \psi^{(4)}=(\chi_0^2+\chi_1\chi_0'+\chi_0(\chi_1^2+2\chi_1')+\chi_0'')\psi+
 (\chi_1^3+2\chi_0'+\chi_1(2\chi_0+3\chi_1')+\chi_1'')\psi'.
$$
With the help of (\ref{u1}) and the last equality we rewrite $L_4\psi=z\psi$ in the form
$$
P_1\psi+P_2\psi'=z\psi,
$$
where
$$
 P_1=f_0+f_2\chi_0+\chi_0^2+\chi_1\chi_0'+\chi_0(\chi_1^2+2\chi_1')+\chi_0'',
$$
$$
 P_2=f_1+f_2\chi_1+\chi_1^3+2\chi_0'+\chi_1(2\chi_0+3\chi_1')+\chi_1''.
$$
This gives
\begin{equation}\label{e11}
 P_1=z=\frac{1}{k^2},\qquad \ P_2=0.
\end{equation}
From (\ref{u3}) we have
$$
 P_1-\frac{1}{k^2}=\frac{f_2+2a_0}{k}+(f_0+a_0(f_2+a_0)+2(a_1+b_1')+a_0'')+O(k)=0,
$$
$$
 P_2=(f_1+2(b_1+a_0'))+O(k)=0.
$$
From here we find the coefficients of $L_4$.

Operator $L_4$ is self-adjoint if $f_1=f_2'$, i.e. at $b_1=0$. Lemma 1 is proved.

\vspace{0.4cm}

If $\chi_1$ satisfies (\ref{u2})
then
$
 \chi_1=\sum_{s>1}b_{2s}k^{2s},
$
hence, by Lemma 1 $L_4$ is self-adjoint.

\vspace{0.4cm}

Let us prove the inverse part of Theorem 1. We assume that $L_4$ is self-adjoint
$$
 L_4=L_4^*=\partial_x^4+f_2(x)\partial_x^2+f_2'(x)\partial_x+f_0(x).
$$
If $\psi_1,\psi_2\in Ker(L_4-z)$, then
$$
 \psi_1L_4\psi_2-\psi_2L_4\psi_1=\partial_x(\psi_1\psi_2'''-\psi_2\psi_1'''-(\psi_1'\psi_2''-\psi_2'\psi_1'')+f_2(\psi_1\psi_2'-\psi_2\psi_1'))=0.
$$
Hence, on the space $Ker(L_4-z)$ the following skew-symmetric bilinear form
$$
 (.,.):Ker(L_4-z)\times Ker(L_4-z)\rightarrow {\mathbb C},
$$
$$
 (\psi_1,\psi_2)=\psi_1\psi_2'''-\psi_2\psi_1'''-(\psi_1'\psi_2''-\psi_2'\psi_1'')+f_2(\psi_1\psi_2'-\psi_2\psi_1')
$$
is defined. Let $\psi_1(x,P),\psi_2(x,P)$ satisfy the equation (\ref{u1}).
Using
$$
 \psi_i'''=(\chi_0+\chi_1^2+\chi_1')\psi_i'+(\chi_0\chi_1+\chi_0')\psi_i
$$
we get
$$
 (\psi_1,\psi_2)=(\psi_1\psi_2'-\psi_2\psi_1')(f_2+2\chi_0+\chi_2^2+\chi_1').
$$
Since $\psi_1,\psi_2$ satisfy the second order differential equation (\ref{u1}) we have,
$$
 (\psi_1,\psi_2)=e^{\int\chi_1(x,z,w)dx}g_1(z,w)\left(f_2(x)+2\chi_0(x,z,w)+\chi_1^2(x,z,w)+\chi_1'(x,z,w)\right)
$$
$$
 =g_2(z,w),
$$
where $g_1(z,w),g_2(z,w)$ are some functions on $\Gamma$.
Let us represent $\chi_1$ in the form
$$
 \chi_1(x,z,w)=G_1(x,z)+wG_2(x,z),
$$
where $G_1,G_2$ are rational functions on $\Gamma$. Let
$$
 \tilde{G}_1(x,z)=\int G_1(x,z)dx,\qquad \tilde{G}_2(x,z)=\int G_2(x,z)dx,
$$
then
$$
 e^{\tilde{G}_1(x,z)}\left(e^{\tilde{G}_2(x,z)}\right)^w\frac{g_1(z,w)}{g_2(z,w)}=\frac{1}{f_2+2\chi_0+\chi_2^2+\chi_1'}.
$$
From the last identity it follows that for arbitrary $x=x_1, x=x_2$ the function
$$
 e^{\tilde{G}_1(x_1,z)-\tilde{G}_1(x_2,z)}\left(e^{\tilde{G}_2(x_1,z)-\tilde{G}_2(x_2,z)}\right)^w
$$
is a rational function on $\Gamma$. This is possible only if
$$
 \tilde{G}_2(x_1,z)-\tilde{G}_2(x_2,z)=0,
$$
or equivalent $G_2=0$. Hence, $\chi_1=G_1(x,z)$. This means that $\chi_1$ is invariant under the involution $\sigma$.
Thus, Theorem 1 is proved.

\section{Proof of Theorem 2}

Assume that $\chi_1$ is invariant under $\sigma$, then by (\ref{u3})--(\ref{u5}) we have
$$
 \chi_0=\frac{H_1(x)}{z-\gamma_1(x)}+\dots+
 \frac{H_g(x)}{z-\gamma_g(x)}+\frac{w(z)}{(z-\gamma_1(x))\dots(z-\gamma_g(x))}+\kappa(x),
$$
$$
 \chi_1(x,P)=-\frac{\gamma'_1(x)}{z-\gamma_1(x)}-\dots-
 \frac{\gamma'_g(x)}{z-\gamma_g(x)},
$$
where $H_i(x),\kappa(x)$ are some functions. In the neighbourhood of $q$ the function $\chi_0$ has the expansion
$$
\chi_0=\frac{1}{k}+\kappa+\left(\gamma_1+\dots+\gamma_g+\frac{c_{2g}}{2}\right)k+O(k^2).
$$
Hence, by Lemma 1
\begin{equation}\label{u8}
 V=-\kappa,\qquad W=-2(\gamma_1+\dots+\gamma_g)-c_{2g}.
\end{equation}
Thus
$$
 \chi_0=\frac{Q_1}{Q}+\frac{w}{Q}-V(x),\qquad \chi_1(x,P)=\frac{Q'}{Q}.
$$
Let us substitute $\chi_0,\chi_1$ into (\ref{e11}). From $P_2=0$ we get $Q_1=-\frac{Q''+s}{2}$, where $s$ is a constant. From $P_1=z$ we get
$$
 s^2-4sw+4w^2-4(z-W)Q^2+4V(Q')^2-(Q'')^2+2Q'Q^{(3)}
$$
$$
 -2Q(2V'Q'+4VQ''+Q^{(4)})=0.
$$
The last identity is possible only if $s=0$ because $Q$ is a polynomial in $z$. Theorem~2 is proved.

Let us differentiate (\ref{e1}) in $x$ and divide the result by $Q$. We get the following equation.

\vspace{0.4cm}

\noindent{\bf Corollary 1}  {\it The functions $Q,W,V$ satisfy the equation
$$
 Q^{(5)}+4VQ^{3}+2Q'(2z-2W-V'')+6V'Q''-2QW'=0.
$$}

Let us substitute $z=\gamma_j$ in (\ref{e1}). It gives
$$
 V(x)=\left(\frac{(Q'')^2-2Q'Q^{(3)}-4F_g(z)}{4(Q')^2}\right)\mid_{z=\gamma_j}.
$$
We get $g-1$ equations on $\gamma_1(x),\dots,\gamma_g(x)$.

\vspace{0.4cm}

\noindent{\bf Corollary 2} {} {\it The functions $\gamma_1(x),\dots,\gamma_g(x)$ satisfy the equations
$$
 \left(\frac{(Q'')^2-2Q'Q^{(3)}-4F_g(z)}{4(Q')^2}\right)\mid_{z=\gamma_j}=\left(\frac{(Q'')^2-2Q'Q^{(3)}-4F_g(z)}{4(Q')^2}\right)\mid_{z=\gamma_k}.
$$}

\vspace{0.4cm}

\section{Proof of Theorem 3}

 Let
\begin{equation}\label{v6}
 \chi_0=-\frac{1}{2}\frac{Q''}{Q}+\frac{\sqrt{F_{g}(z)}}{Q}-(\alpha_3x^3+\alpha_2x^2+\alpha_1x+\alpha_0),
\end{equation}
\begin{equation}\label{e20}
 \chi_1=\frac{Q'}{Q}.
\end{equation}
Let us consider the equations (\ref{e1}) where $V,W$ are potentials of the operator $L^{^{\sharp}}_4$
$$
 4F_g(z)=4(z-g(g+1)\alpha_3x)Q^2-4(\alpha_3x^3+\alpha_2x^2+\alpha_1x+\alpha_0)(Q')^2+(Q'')^2-2Q'Q^{(3)}
$$
\begin{equation}\label{e21}
 +2Q(2(3\alpha_3x^2+2\alpha_2x+\alpha_1)Q'+4(\alpha_3x^3+\alpha_2x^2+\alpha_1x+\alpha_0)Q''+Q^{(4)}).
\end{equation}

We prove that the nonlinear equation (\ref{e21}) has a polynomial solution $Q(x,z)$
of degree $g$ in $z$ and degree $g$ in $x$ for some polynomial $F_g(z)$.
After that we prove that $\chi_0,\chi_1$ satisfy (\ref{u6}) for
the curve $w^2=F_g(z)$. The functions $\chi_0,\chi_1$ have required asymptotic (\ref{u3}) in $q=\infty$.
From here it follows that $L_4^{^{\sharp}}$ commutes with an operator of order $4g+2$ corresponding to the rational function $w$
on $\Gamma$ with the unique pole of order $2g+1$ at $q$.

\vspace{0.4cm}

\noindent{\bf Lemma 2} {\it Equation (\ref{e21}) has a solution of the form
\begin{equation}\label{e22}
 Q=(z-\gamma_1(x))\dots(z-\gamma_g(x)),
\end{equation}
for some polynomial $F_g(z)$ of degree $2g+1$.}

\vspace{0.4cm}

\noindent{\bf Proof.} {} Let us differentiate both sides of (\ref{e21}) with respect
to $x$ and divide the result by $Q$
$$
 Q^{(5)}+4(\alpha_3x^3+\alpha_2x^2+\alpha_1x+\alpha_0)Q^{(3)}+4(\alpha_2-(g^2+g-3)\alpha_3x+z)Q'
$$
\begin{equation}\label{e23}
 +6(3\alpha_3x^2+2\alpha_2x+\alpha_1)Q''-2g(g+1)\alpha_3Q=0.
\end{equation}
We find a solution of (\ref{e23}) as a polynomial in $x$
\begin{equation}\label{e24}
 Q=\delta_gx^g+\dots+\delta_1x+\delta_0,\qquad \delta_i=\delta_i(z).
\end{equation}
From (\ref{e23}) we have
$$
 \delta_s=\frac{(s+1)}{\alpha_3(g-s)(s+g+1)(2s+1)}\left(2(\alpha_2(s+1)^2+z)\delta_{s+1}+\alpha_1(s+2)(2s+3)\delta_{s+2}\right.
$$
\begin{equation}\label{e25}
 \left.+2\alpha_0(s+2)(s+3)\delta_{s+3}+1/2(s+2)(s+3)(s+4)(s+5)\delta_{s+5}\right),
\end{equation}
where $0\leq s<g-1$, $\delta_g$ is a constant, and $\delta_s=0$ at $s>g$. In particular
\begin{equation}\label{e26}
 \delta_{g-1}=\frac{\delta_g(\alpha_2g^2+z)}{\alpha_3(2g-1)}.
\end{equation}
From (\ref{e25}) it follows that $Q$ is a polynomial of
degree $g$ in $z$, and up to the multiplication by a constant, the polynomial $Q$ has the form (\ref{e22}).
The right-hand side of (\ref{e21}) has degree $2g+1$.
 Lemma 2 is proved.

\vspace{0.4cm}

\noindent{\bf Lemma 3} {\it {} The polynomial $Q$ has no multiple root in $z$
$$
 \gamma_i\ne\gamma_j\ \mbox{at}\ i\ne j.
$$}

\vspace{0.4cm}

\noindent{\bf Proof.} {} Let us represent $Q$ in the form
$$
 Q=Q_H+\tilde{Q},
$$
where $Q_H$ is a homogeneous polynomial in $x,z$
$$
 Q_H=\tilde{\delta}_gx^g+\tilde{\delta}_{g-1}x^{g-1}z+\tilde{\delta}_{g-2}x^{g-2}z^2+\dots+\tilde{\delta}_0z^g, \qquad \tilde{\delta}_0,\tilde{\delta}_g\ne0
$$
and ${\rm deg}\tilde{Q}<g$. Since $\tilde{\delta}_g\ne0$, the polynomial $Q$ has no constant roots (i.e. $\gamma_i\ne const$).

Let us note that $Q$ has no multiple roots of order higher than 2. Indeed, if $Q=(z-\gamma_i(x))^p\hat{Q},\ p>2$, then from (\ref{e21})
$F_g(\gamma_i(x))=0$, but this is impossible.

If $Q$ has multiple roots, then $Q_H$ also has multiple roots. This follows from the following fact.
The discriminant of $Q$ is a polynomial $b_Nx^N+b_{N-1}x^{N-1}+\dots+b_0$ in $x$. The discriminant of $Q_H$ is $b_Nx^N$, so if the discriminant of
$Q$ is equal to zero, then the discriminant of $Q_H$ is also zero.

From (\ref{e25}) it follows that
$$
 \tilde{\delta}_s=\frac{2(s+1)\tilde{\delta}_{s+1}}{\alpha_3(g-s)(s+g+1)(2s+1)},\qquad 0\leq s\leq g-1,
$$
and that $Q_H$ satisfies the equation
$$
 2\alpha_3x^3Q_H^{(3)}+2((3-g-g^2)\alpha_3x+z)Q_H'+9\alpha_3x^2Q_H''-g(g+1)\alpha_3Q_H=0.
$$
Let us multiply this equation by  $Q_H$ and integrate in $x$. We get
$$
 \tilde{F}_g(z)+(g(g+1)\alpha_3x-z)Q_H^2+\alpha_3x^3(Q_H')^2-\alpha_3x^2Q_H(3Q_H'+2xQ_H''))=0,
$$
where $\tilde{F}_g(z)$ is a polynomial of degree $2g+1$ in $z$.

 From the last equation it follows that if $Q_H$ has multiple roots, then the polynomial $\tilde{F}_g(z)$ has the same roots. However, this is impossible,
because all roots of $\tilde{F}_g(z)$ are constant, but $Q_H$ has not constant roots.
Lemma 3 is proved.

\vspace{0.4cm}

\noindent{\bf Lemma 4} {\it {} If $(\alpha_0,\dots,\alpha_3)\in U$, the curve $w^2=F_g(z)$ is
nonsingular, where $U\subset{\mathbb C}^4$ is some Zariski open set.
}

\vspace{0.4cm}

\noindent{\bf Proof.} {} The idea of the proof is the following. We represent $F_g$ in the form
$$
 F_g(z)=F_g^0(z)+\alpha_3F_g^1(z)+O(\alpha_3^2),
$$
and prove that $F_g^0(z)+\alpha_3F_g^1(z)$ has not multiple roots. Therefore, $F_g(z)$ has not multiple roots for small $\alpha_3$, and consequently
for $(\alpha_0,\dots, \alpha_3)\in U$.

Let us consider (\ref{e24})--(\ref{e26}). We put $\delta_g=\alpha_3^g$, then
$$
 \delta_{g-1}=\alpha_3^{g-1}\frac{\alpha_2g^2+z}{2g-1}.
$$
Moreover, from (\ref{e24}) it follows that $Q$ has the form
\begin{equation}\label{e27}
 Q=\alpha_3^gx^g+\dots+
 \alpha_3^sx^s(p_{s}(z)+\alpha_3q_{s}(z)+O(\alpha_3^2))+\dots+(p_{0}(z)+\alpha_3q_{0}(z)+O(\alpha_3^2)).
\end{equation}
 Let us note that from (\ref{e26}) it follows that
$$
 p_g=1,\qquad p_{g-1}=\frac{\alpha_2g^2+z}{2g-1},\qquad q_g=0,\qquad q_{g-1}=0.
$$
Let us substitute (\ref{e27}) into (\ref{e21}). We get
$$
 F_g(z)=p_0^2(z)z+\alpha_3p_0(z)(\alpha_1p_1(z)+2q_0(z)z)+O(\alpha_3^2),
$$
so,
$$
 F_g^0(z)=p_0^2(z)z,\qquad F_g^1(z)=p_0(z)(\alpha_1p_1(z)+2q_0(z)z).
$$
To prove Lemma 4 it is enough to prove that
$p_0(z)z$ and $\alpha_1p_1(z)+2q_0(z)z$ have no common roots.

Let us find $p_i$ and $q_i$. For this we again substitute (\ref{e27}) into (\ref{e23}) and find the coefficients at $\alpha_3^{i+1}x^i$ and $\alpha_3^{i+2}x^i$.
These coefficients must be equal to zero. It gives us
\begin{equation}\label{e28}
 p_i=\frac{2(i+1)(\alpha_2(i+1)^2+z)}{(2i+1)(g^2+g-i^2-i)}p_{i+1},\qquad 0\leq i\leq g-1,
\end{equation}
\begin{equation}\label{e29}
  q_i=\frac{2(i+1)(\alpha_2(i+1)^2+z)}{(g-i)(g+i+1)(2i+1)}q_{i+1}+\frac{\alpha_1(i+1)(i+2)(2i+3)}{(g-i)(g+i+1)(2i+1)}p_{i+2},
\end{equation}
where $0\leq i\leq g-2.$ Hence
$$
 p_i(z)=(\alpha_2(i+1)^2+z)\dots(\alpha_2g^2+z)A_i,\qquad 0\leq i\leq g-1,
$$
where $A_i$ is a constant. Thus to prove that $p_0(z)z$ and $\alpha_1p_1(z)+2q_0(z)z$ have no common roots we should prove that
$z=-\alpha_22^2,\dots,z=-\alpha_2g^2$ are not roots of $q_0(z)$. Assume that $q_0(-\alpha_2s^2)=0$ for some $s$, {} $2\leq s\leq g.$
 From (\ref{e28}) it follows that $p_k(-\alpha_2s^2)=0$ at $0\leq k<s$, $p_k(-\alpha_2s^2)\ne 0$ at $k\geq s$, and from (\ref{e29}) it
follows that $q_k(-\alpha_2s^2)=0$ at $0\leq k\leq s-2$.

First of all we consider the case $s=g$. If $i=g-2$, then (\ref{e29}) yields
$$
  q_{g-2}(z)=\frac{\alpha_1(g-1)g(2s-1)}{2(2g-1)(2g-3)}p_g(z).
$$
Hence, if $q_0(-\alpha_2g^2)=0$, then $q_{g-2}(-\alpha_2g^2)=0$, but this is impossible, since $p_g=1$, so  $s<g.$

Formulas (\ref{e28}), (\ref{e29}) at $i=s-2, i=s-1$ give us
$$
q_{s-2}-\frac{2(s-1)(\alpha_2(s-1)^2+z)}{(g-s+2)(g+s-1)(2s-3)}
 \frac{2s(\alpha_2s^2+z)q_s+\alpha_1s(s+1)(2s+1)p_{s+1}}{(g-s+1)(g+s)(2s-1)}-
$$
$$
 \frac{\alpha_1(s-1)s(2s-1)}{(g-s+2)(g+s-1)(2s-3)}\frac{2(s+1)(\alpha_2(s+1)^2+z)}{(2s+1)(g^2+g-s^2-s)}p_{s+1}=0.
$$
Let $z$ be $-\alpha_2s^2$.
After the simplification we have
$$g^2+g-3s^2=0.$$
This is impossible, hence $q_0(-\alpha_2s^2)\ne0$ and $F_g^0(z)+\alpha_3F_g^2(z)$ has no multiple roots.
Lemma 4 is proved.

Functions $\chi_0,\chi_1$ are rational functions on the curve $w^2=F_g(z)$. Let $k=\frac{1}{\sqrt{z}}$ be a local parameter near $q=\infty$. Functions
$\chi_0,\chi_1$ have asymptotic (\ref{u2}). By Lemma, 3 $\chi_0$ and $\chi_1$ have simple poles $P_i^{\pm}=(\gamma_i,\pm\sqrt{F_g(\gamma_i)}).$
Let us choose in the neighbourhood of $P_i^{\pm}$ the local parameter $z-\gamma_i(x)$.

\vspace{0.4cm}

\noindent{\bf Lemma 5} {\it {} Functions $\chi_0,\chi_1$ satisfy the equation (\ref{u6})}.

\vspace{0.4cm}

\noindent{\bf Proof.} From (\ref{e20}) we have
$$
\chi_1(x,P)=\frac{-\gamma'_i(x)}{z-\gamma_i(x)}+d_{i,1}(x)+O(z-\gamma_i(x))
$$
for some $d_{i,1}(x).$ Function $\chi_1$ has simple poles at $\gamma_i(x)$, thus
$$
\chi_0(x,P)=\frac{-v_{i,0}(x)\gamma'_i(x)}{z-\gamma_i(x)}+d_{i,0}(x)+O(z-\gamma_i(x)),
$$
for some $v_{i,0}(x),d_{i,0}(x)$. By our construction $\chi_0$, $\chi_1$ satisfy (\ref{e11}). Let us substitute
$\chi_0$, $\chi_1$ in (\ref{e11}). We get
$$
 \frac{(v_{i,0}^2(x)-d_{i,0}(x)+d_{i,1}(x)v_i(x)-v_i'(x))(\gamma_i'(x))^2}{(z-\gamma_i(x))^2}+O\left(\frac{1}{z-\gamma_i(x)}\right)=0.
$$
Hence $d_{i,0}(x), d_{i,1}(x), v_{i,0}(x)$ satisfy (\ref{u6}).
Lemma 5 and Theorem 3 are proved.

\vspace{0.4cm}

Operator $L^{^{\sharp}}_{4g+2}$ commuting with $L_4^{^{\sharp}}$ can be found from
$L_4^{^{\sharp}}L^{^{\sharp}}_{4g+2}=L^{^{\sharp}}_{4g+2}L_4^{^{\sharp}}$. For the simplicity of the formulas we restrict ourselves to the case
$\alpha_1=\alpha_2=0,\alpha_3=1$. Let us introduce the notations:
{} $H=\partial_x^2+x^3+\alpha_0,$ {} $\langle A,B\rangle=AB+BA.$

\vspace{0.4cm}

\noindent{\bf Examples.}
\vspace{0.2cm}

\noindent a) $g=2:$
$$
 L^{^{\sharp}}_{10}=H^5+\frac{15}{2}\langle x,H^3\rangle+45\langle x^2,H\rangle,
$$
$$
 F_2(z)=z^5+27\alpha_0 z^2+81.
$$
\noindent b) $g=3:$
$$
 L^{^{\sharp}}_{14}=H^7+21\langle x,H^5\rangle+\frac{945}{2}\langle x^2,H^3\rangle-
 5418H^2
 +\frac{45}{2}\langle 113\alpha_0+287x^3,H\rangle-486x,
$$
$$
 F_3=z^7+594\alpha_0z^4-2025z^2+91125\alpha_0^2z.
$$

\end{document}